\documentclass[twocolumn]{article}  

\usepackage{natbib}
\bibpunct{(}{)}{;}{a}{}{,} 

\usepackage{float}
\usepackage{multirow}

\usepackage{booktabs}
\usepackage{graphicx}
\usepackage{adjustbox} 
\usepackage{amsmath}    
\usepackage{newtxtext}  
\usepackage{newtxmath}  
\usepackage{soul}
\usepackage{authblk} 
\usepackage{color}
\definecolor{pinky}{rgb}{0.4, 0.,0.83}

\usepackage[colorlinks=true,
            linkcolor=blue,
            citecolor=blue,
            urlcolor=blue]{hyperref}

\begin{document}

   \title{A New Multi-Constraint Potential Field Source Surface (PFSS) Extrapolation Model}

\author[1]{C. Antonio}
\author[2]{I. Chifu}
\author[1,3]{R. Gafeira}
\author[4,5]{J.J.G. Lima}

\affil[1]{Instituto de Astrofísica e Ciências do Espaço, Department of Physics, University of Coimbra, PT3040-004 Coimbra, Portugal}
\affil[2]{Institute for Astrophysics and Geophysics, Friedrich-Hund-Platz 1, 37077, Göttingen, Germany}
\affil[3]{Geophysical and Astronomical Observatory, Faculty of Science and Technology, University of Coimbra, Rua do Observatório s/n, PT3040-004 Coimbra, Portugal}
\affil[4]{Departamento de Física e Astronomia, Faculdade de Ciências, Universidade do Porto, Rua do Campo Alegre 687, 4169-007 Porto, Portugal}
\affil[5]{Instituto de Astrofísica e Ciências do Espaço, Universidade do Porto, CAUP, Rua das Estrelas, PT4150-762 Porto, Portugal}

\affil[ ]{\textit{carlos.antonio@student.uc.pt}}

\twocolumn[
  \begin{@twocolumnfalse}
    \maketitle

\begin{abstract}   The Potential Field Source Surface (PFSS) model remains the most widely used method for extrapolating the global coronal magnetic field. Given the current spatial resolution of the full Sun, PFSS performs well at large scales and is computationally efficient. The PFSS approach, however, fails when electric currents distort the coronal field from its potential state. More advanced extrapolation techniques, such as nonlinear force-free field (NLFFF) models, can capture non-potential effects but are significantly more computationally demanding for global studies. Observational techniques allow the reconstruction of the three-dimensional geometry of coronal loops, which trace the magnetic field in the corona.

   The goal of this work is to develop a new implementation to multi-constrain the global PFSS model by incorporating three-dimensional coronal loop information, thereby improving its consistency with observations while preserving its computational efficiency. Although the model remains constrained by the PFSS field derived from photospheric observations, it allows the magnetic field to deviate from the potential, minimum-energy state within the loop influence regions, while maintaining control over its divergence and force-freeness.
   We adapted the NLFFF optimization formalism to the PFSS framework, enabling the inclusion of multiple physical constraints. Up to three terms were considered in the functional: a divergence-free term, a loop term, and a force-free term. The newly developed second-order finite-difference Python algorithm was tested with synthetic coronal loop data, using Carrington rotation 2284 as the lower boundary condition.
   The method produces magnetic field solutions that are more consistent with the geometry of the included coronal loops, controlling the divergence and force-freeness levels associated with the coronal loop inclusion.
   Our method shows that 3D coronal loop information can be consistently incorporated into the PFSS, while largely preserving its computational efficiency, even when a substantial number of loops is considered. 
   
\vspace{0.5cm} 
 \end{abstract}
\end{@twocolumnfalse}
]

%

\section{Introduction}

The coronal magnetic field (CMF) is fundamental for understanding the physical processes and structures in the solar corona. Unlike for the photosphere, routine CMF measurements are not yet available. Such measurements require full 3D information from a low-density, optically thin medium \citep[e.g.]{Wiegelmann2015}. The Zeeman effect, typically used to measure the photospheric magnetic field, is not as effective in the corona due to the small splitting caused by the weaker magnetic field \citep[e.g.]{Yang2024, Wiegelmann2021}.
In addition, coronal spectral lines are faint and strongly broadened due to the low density and high temperatures \citep[e.g.][]{Lin2004, Cargill2008}, further increasing the measurement uncertainties. Several attempts have been made to measure the coronal magnetic field \citep[e.g.][]{Lin2000ApJ...541L..83L}, and in some specific regions it has been successfully measured using radio emission diagnostics \citep{White1999, Gibson2016}.\par

To obtain the CMF, one typically relies on magnetic field extrapolations, the process of reconstructing the magnetic field measured at the solar surface \citep[e.g.][]{Wiegelmann2021}. The most common extrapolation method is the potential field source surface (PFSS) \citep[e.g.][]{Altschuler1969, Schatten1969}. Potential fields are unique minimal energy solutions that do not contain free energy. PFSS is widely used today for global studies, benefiting from its computational efficiency. In the PFSS method, the source surface is usually placed at 2.5 R$_\odot$ \citep[e.g.][]{Riley2006}, on which the field lines are forced to be purely radial, mimicking the effect of the solar wind. In the coronal volume, the magnetic field is obtained by solving the Laplace equation using the observed vertical component of the photospheric field as a lower boundary condition. While potential fields and the PFSS method are good at recovering larger-scale field topology \citep{Aulanier2005}, these sometimes fail when free energy is significant and when strong electric currents flow in the corona and distort the magnetic field, namely within active regions \citep[e.g.][]{Wiegelmann2004}. \par
The state of the art of force-free extrapolation modeling lies in the realm of NLFFFs, which are able to account for free energy effects. Several NLFFF approaches have been developed \citep[e.g.][]{Wiegelmann2021}. Among them, the NLFFF-optimization method \citep[e.g.][]{Wheatland2000} is one of the best performing ones for active regions for which reliable data exist \citep[e.g.][]{Schrijver2006, Yeates2018}. The former uses vector magnetograms as boundary conditions. 
The information regarding the shape of coronal loops, which are bright, arch-like plasma structures, is typically obtained in extreme ultraviolet (EUV) emission images of the solar corona. Their 3D geometry can be reconstructed using stereoscopic techniques that combine emission images from different viewing angles \citep[e.g.][]{Wiegelmann-Inhester2006SoPh..236...25W, Feng2007, Aschwanden2011}. Because the coronal plasma is highly conductive, these loops trace the geometry of the magnetic field under the frozen-in condition, allowing their 3D structure to be used as a proxy for the coronal magnetic field 3D geometry \citep[e.g.][]{Wiegelmann2015}. Coronal loop geometries have been used as observational constraints in purely NLFFF modeling, mitigating inconsistencies between the NLFFF assumption and the lower boundary \citep{Chifu2015, Chifu2017}, and potentially reducing the impact of finite plasma-$\beta$ effects \citep{Peter2015}. Such information is incorporated as observational constraints to increase the magnetic field's tangency to coronal loops whose 3D shape is known from observations. \par

While NLFFF's extrapolations are more accurate than PFSS and account for free energy, the former is much slower than the PFSS for global field studies. On the other hand, observationally constraining the magnetic field, namely improving its orientation relative to the observed one, does not necessarily lead to smaller free energy solutions \citep{Peter2015}. The magnetic field computed using this type of approach may be closer to the real one and not necessarily associated with energy minima configurations given a lower boundary \citep{Peter2015}. With the increasing availability of different coronal observations, the inclusion of the observational constraints in the magnetic field modeling is one of the natural approaches to pursue \citep[e.g.][]{Wiegelmann2015, Peter2015}.\par

The main goal of this work is to build a new multi-constrained PFSS model that allows for the inclusion of the 3D coronal loop geometry, allowing the magnetic field to be more tangent to the loops included, while we assess and control the divergence or even the degree of force-freeness, two additional constraints included for self-consistency. By doing so, we correct the PFSS field and improve it at the coronal level. The model allows the inclusion of a significant number of loops and retains the most relevant PFSS advantage, which is its computational efficiency. As a result, by incorporating the 3D loop geometry, we obtain a more realistic description of the coronal magnetic field compared with the classic PFSS. In this paper, we describe and analyze two different versions of the code and their results. \par
The paper is structured as follows: in section \ref{multi-constrained-model-method-section}, we describe the new proposed multi-constrained PFSS model along with a description of its implementation and algorithm. In section \ref{Data} we describe the synthetic data used for testing, and in section \ref{results-section} we show the results from the different test runs and configurations, along with corresponding analysis. In section \ref{conclusions} we present the conclusions from this paper.

\section{The Multi-Constrained PFSS Model}\label{multi-constrained-model-method-section}

The inclusion of the observed 3D coronal loop geometry must be performed in a way that ensures that the magnetic field in the computational box remains divergence-free. This requires an incremental approach, in which the information derived from the coronal loop shapes is gradually incorporated as the magnetic field evolves, while remaining consistently constrained by the photospheric observations. To multi-constrain the PFSS model, we adopt the mathematical formulation of the NLFFF optimization method \citep{Wiegelmann2004} and adapt it to the present context. The optimization approach involves defining a given functional, composed of a number of positive terms, each of which quantifies constraints that the final solution must satisfy \citep[e.g.][]{Chifu2015}. The fulfillment of those constraints, expressed as terms in the functional, is done by means of an optimization procedure, in which these are minimized simultaneously and iteratively. The reason behind the choice of this model among others is its flexibility when it comes to change, including or excluding constraints \citep[e.g.][]{Wiegelmann2010, Chifu2015}. We can define a given total functional ($L_t$) as a sum of different constraints expressed in the terms $L_t=L_1+L_2+(\ldots)+L_n$. \par
The divergence term is an integral over the volume $V$, as defined in \cite{Wiegelmann2004}, which in spherical coordinates is written as

\begin{equation}\label{div-term}
L_1=\int_V\omega(r, \theta, \phi)|\nabla \cdot \boldsymbol{B}|^2 r^2 \sin \theta \mathrm{~d} r \mathrm{~d} \theta \mathrm{~d} \phi
\end{equation}
To observationally constrain the model in the corona, we include the $L_2$ term which is related to the tangency between the magnetic field and the loops. This is a line integral along each loop included 

\begin{equation}\label{tangency-term}
L_2=\sum_i \frac{1}{\int_{\mathbf{c}_i} d s} \int_{\mathbf{c}_i} \frac{\left|\boldsymbol{B} \times {t}_i\right|^2}{\sigma_{c_i}^2} d s
\end{equation}
where 

\begin{equation}
t_i=\frac{d \mathbf{c}_i}{d s}
\end{equation}
are the tangent vectors of a given loop obtained from the observations. The integral term in the denominator of the expression \ref{tangency-term}, $\int_{\mathbf{c}_i} d s_i$, is the length of the \textit{i}th loop, along which integration occurs. The summation symbol in expression \ref{tangency-term} is a sum over all the loops included, meaning this term increases with the number of loops one includes. The loop term \ref{tangency-term} also contains the estimated loop errors, represented by the $\sigma_{c_i}^2$ term, which in the absence of real data are set to one. \par

 Including a force-free constraint will allow us to assess and constrain the force-freeness of the output magnetic field, preserving the global force-free nature of the field. We make use of the force-free expression introduced by \cite{Wiegelmann2004}, which in spherical coordinates has the following shape

 \begin{equation}\label{ff-term}
L_3=\int_V \omega(r, \theta, \phi)B^{-2}|(\nabla \times \boldsymbol{B}) \times \boldsymbol{B}|^2 r^2 \sin \theta \mathrm{~d} r \mathrm{~d} \theta \mathrm{~d} \phi
\end{equation}

The force-free constraint is typically defined as the first term (i.e., $L_1$) of the functional as in \cite{Chifu2015}, or minimized together with the solenoidal constraint forming a volume term $L$ originally in \cite{Wiegelmann2004}. Since we use different setups in which we include or exclude different terms of the functional, we adopt a distinct nomenclature $L_n$.\par

For the success of functional minimization, one must compute the necessary, incremental, field adjustments required for $dL_{t}/dt$. In practice, derivatives of each constraint term, $dL_{n}/dt$, are computed according to the procedure described in \cite{Wheatland2000} and \cite{Wiegelmann2004}. By solving the pseudo-time evolution as
\begin{equation}
\frac{\partial \boldsymbol{B}}{\partial t}=\mu_nF_n,
\end{equation}
and by imposing fixed boundary conditions at both the photospheric surface and the source surface, we ensure $dL_{\mathrm{n}}/dt<0$, if the $\mu_n$ are positive and sufficiently small. Each $F_n$ represents the functional derivative of the corresponding constraint term $L_n$ with respect to the magnetic field. The source surface, along with the base of the computational domain, remains fixed, ensuring not only that the field is equal to the field derived from photospheric observations, but also avoiding flux removal/injection through the boundaries. This constantly enforces the observed photospheric field constraint, ensuring the observed photospheric field constrains the computed magnetic field above at every iteration, and consequently the final solution. The boundary conditions are satisfied by ensuring the weights $\omega(r, \theta, \phi)$ in Eqs. \ref{div-term} and \ref{ff-term} decay to zero at both boundaries, following a cosine profile for a smoother boundary prescription. The cosine decaying profile has been used in the NLFFF context, both in Cartesian \citep{Wiegelmann2004} and spherical coordinates \citep{Tadesse2009, Tadesse2011}, for the top and lateral boundaries (when applicable), which has been acknowledged as the best performing decaying function. Since we are interested in a global PFSS, lateral boundaries are absent, and contrary to previous works, we also use it on the bottom boundary. Using this procedure, the three components of the magnetic field on the base and top of the computational domain remain fixed. \par

After some vector identities, $dL_{\mathrm{2}}/dt$ can be formulated as

\begin{equation}
\frac{1}{2} \frac{d L_2}{d t}=-\sum_i \frac{1}{d s_i} \int_{\mathrm{c}_i} \frac{\partial \boldsymbol{B}}{\partial t} \cdot F_2 d s
\end{equation}

where $F_2=-\left[\boldsymbol{B}\left|t_i\right|^2-t_i\left(\boldsymbol{B} \cdot t_i\right)\right]$, is the component of the magnetic field perpendicular to the loop direction, obtained by subtracting from the total field its projection along the loop tangent $t_i$, which satisfies $\left|t_i\right|^2=1$. The derivative $dL_2/dt$ is negative if $\partial \boldsymbol{B}/\partial t=\mu_2 F_2$, indicating that the field evolves to become more tangent to the included loops. One obtains $F_1$ and $F_3$ by analogous procedures, as shown by \cite{Wiegelmann2004} (expression number 15). The small increments $F_1$ and $F_3$ associated with global constraints are applied at every grid point of the computational domain. Constraining the potential field model by the 3D coronal loops, one obtains a new solution of the PFSS for which the magnetic field vector outlines the 3D inserted coronal loops. At the same time, the constraints $F_1$ and $F_3$ dictate how the field must adjust locally to fulfill the global divergence $L_1$ and force-freeness $L_3$ constraints. \par

 We make a terminology distinction between two types of perturbations: (1) "physical" perturbations are associated with the added coronal loops that are capable of disturbing the field from its force-free and potential state; and (2) adjustments, the small numerical corrections $F_n$ that reduce the total functional. The latter is defined in terms of regularization parameters, $\xi_n$, defined as follows \citep{Wiegelmann2004}: 
\begin{center}\label{total-functional-qsi}
\begin{equation}\label{functional-definition}
L_t=\sum_{n=1}^N \xi_n L_n
\end{equation}
\end{center}

where $N$ is the total number of constraints. The coefficients $\xi_n$ are the regularization weights, each of which acts as a scaling factor that controls the relative importance of each competing constraint term in the total functional \ref{functional-definition}. These parameters are set to unity unless stated otherwise. The minimization of \ref{functional-definition} is performed by a Landweber iteration, meaning that at each point the magnetic field becomes

\begin{equation}\label{update}
\boldsymbol{B} \leftarrow \boldsymbol{B}- \sum_{n=1}^N \xi_n \mu_n F_n ,
\end{equation}

where $\mu_n$ is related to the size of $F_n$ adjustments. In Eq. \ref{update}, $\mu_n$ may not necessarily be the same for each constraint one includes, a difference to the \cite{Wheatland2000} and \cite{Wiegelmann2004} formalisms. The initial $\mu_n$'s size must be as large as possible to allow for faster convergence but small enough to allow a monotonic decrease of the functional terms, and to ensure reaching a minimum, and avoiding overshooting following the procedures found as in, for example, \cite{Wiegelmann2004} and \cite{Koumtzis2023}. A $\mu_n$ dynamic update is required for two reasons. First, using the same $\mu_n$ after a given successful iteration does not guarantee a functional decrease in the following iteration. Second, it increases convergence speed. In expression \ref{update}, $\mu_1$, $\mu_3$, and $\mu_2$ are allowed to evolve separately, increasing the efficiency of the model.
Constraining the potential field model by the 3D coronal loops, one obtains a new solution of the PFSS for which the magnetic field vector outlines the 3D inserted coronal loops.  \par

Since different loops are included, with varying lengths and distances to each other, the $n=2$ term of expression \ref{update} is better described by

\begin{equation}\label{mu_2}
\xi_2\mu_2 F_2=\xi_2\sum_{j=1}^m \sum_{i=1}^m \mu_{2, i} \delta_{i j} F_{2,i j} ,
\end{equation}

where $m$ is the total number of loops being included. Expression \ref{mu_2} means $\mu_2$'s are allowed to be adjusted to each loop, reducing the computation time but also the loop term \ref{tangency-term} because different loops may have different $\mu_2$ thresholds. \par

For the equivalent discretized problem, the terms $\mathrm{Ł}_n^{\infty}$ denote the final residual values of the corresponding total functional components after convergence. Typically, the constraints $\mathrm{Ł}_n^{\infty}$ cannot be reduced to zero, as they are bounded by discretization and measurement noise. When several regularization terms are present, the optimal set of parameters $\left(\xi_1, \ldots, \xi_4\right)$ is determined by minimizing $\sum_n \log \mathrm{Ł}_n^{\infty}\left(\xi_1, \ldots, \xi_4\right)$ \citep{10.5555/1805888}. 

The method presented in this work was implemented integrally in \texttt{python} by developing a second-order finite difference algorithm. Since the grid is discrete, constraints $L_n$ must be discretized. The $L_1$ discretization is straightforward, but the $L_2$ assumes the following shape \citep{Chifu2015}

\begin{equation}\label{eq:L2discrete}
L_2 \approx \sum_i \frac{1}{\sum_j \Delta s_j}
      \sum_j \frac{\bigl|\overline{\mathbf{B}}\bigl(c_i(s_j)\bigr)\times \mathbf{t}_i(s_j)\bigr|^{2}}
                   {\sigma_{c_i}^2(s_j)}\,\Delta s_j
\end{equation}

In Eq. \ref{eq:L2discrete}, we discretize the loop parameter $s_j$ (associated with the tangent $t_i(s_j)$ evaluated at the loop point $s_j$ of the \textit{i}th loop), which does not coincide with the grid points where the PFSS is computed. In practice, $\overline{\mathbf{B}}(c_i(s_j))$ used to compute $L_2$ term is the magnetic field interpolated from the grid on the point $s_j$ that belongs to the loop described by \textit{i}th $c_i$ curve. As $L_2$ is a line integral over each loop we include, $F_2$'s are only applied along the latter. The inclusion of the loops will disturb the original field by interpolating $\overline{\boldsymbol{B}}\left(c_i(s)\right)- \xi_2 \mu_{2,i} F_2\left(c_i(s)\right)$, in a given iteration $t$, back to the grid.\par

The $L_3$ force-free constraint requires defining $\epsilon$ for numerical stability, which is set to $10^{-6}$. Therefore 
\begin{equation}\label{eq:L3discrete}
L_3 \approx \sum_{i,j,k}\omega\bigl(r_i,\theta_j,\phi_k\bigr)\,
      \frac{\bigl|(\nabla_h\times\mathbf{B})_{ijk}\times\mathbf{B}_{ijk}\bigr|^{2}}
           {|\mathbf{B}_{ijk}|^{2}+\varepsilon^{2}}\,\Delta V_{ijk}
\end{equation}

where $\Delta V_{i, j, k}=r_i^2 \sin \theta_j \Delta r_i \Delta \theta_j \Delta \phi_k$ and the subscript $h$ means a discrete numerical differential operator. \par

The developed algorithm is summarized into the following steps

\begin{enumerate}
    \item The functional $L_t$ (Eq. \ref{functional-definition}) is computed;
    \item At every $t$, interpolate $\boldsymbol{B}$ on the loops, and an $L_2$ is obtained. $F_2$ adjustments are computed and applied along the latter, yielding a new $L_2$;

    \item The result magnetic field on the loops is interpolated back to the grid;

    \item $F_1$ and $F_3$ are computed and applied at every point of the computational domain;

    \item At the end of each iteration $t$, the code checks the functional value recomputed using the modified field:
    \begin{itemize}
        \item If $L_t(t+dt)<L_t(t)$, the vector field is updated, and $\mu_n$'s are increased ($\mu_n$ is increased by 1.02);

        \item If $L_t(t+d t)>L_t(t)$, there is no update, and $\mu_n$'s are reduced ($\mu_n$ decreases by 2);
    \end{itemize}

    \item Steps 1-5 are iteratively repeated until the functional \ref{functional-definition} does not experience further reduction, becoming stationary within a tolerance of 25 iterations or when a maximum number of 500 iterations is reached (both of which the user is free to change).

\end{enumerate}

The code requires the following specifications:

\begin{itemize}

    \item A 3D position vector ($r,\theta,\phi$) of the loops included, from which it is possible to compute the tangent vectors at each point;
    \item An initial PFSS computed over the grid, which serves as the optimization input field and requires a radial photospheric magnetic field synoptic (or synchronic) map as a boundary condition; 

    \item The PFSS is computed in a spherical grid, regular or irregular, at the center of each grid cell;

    \item Defining the latitudinal range by removing the polar caps to avoid singularities.

\end{itemize}

Up to $2^{\circ}$ were removed to prevent the model from developing singularities at the poles. The effect of this removal is only manifested through the small $F_1$ and $F_3$ adjustments computed over and applied to every point in the computational domain. No parallelization techniques were employed.

\section{Data}\label{Data}

\begin{figure}
    \centering
    \includegraphics[width=1\linewidth]{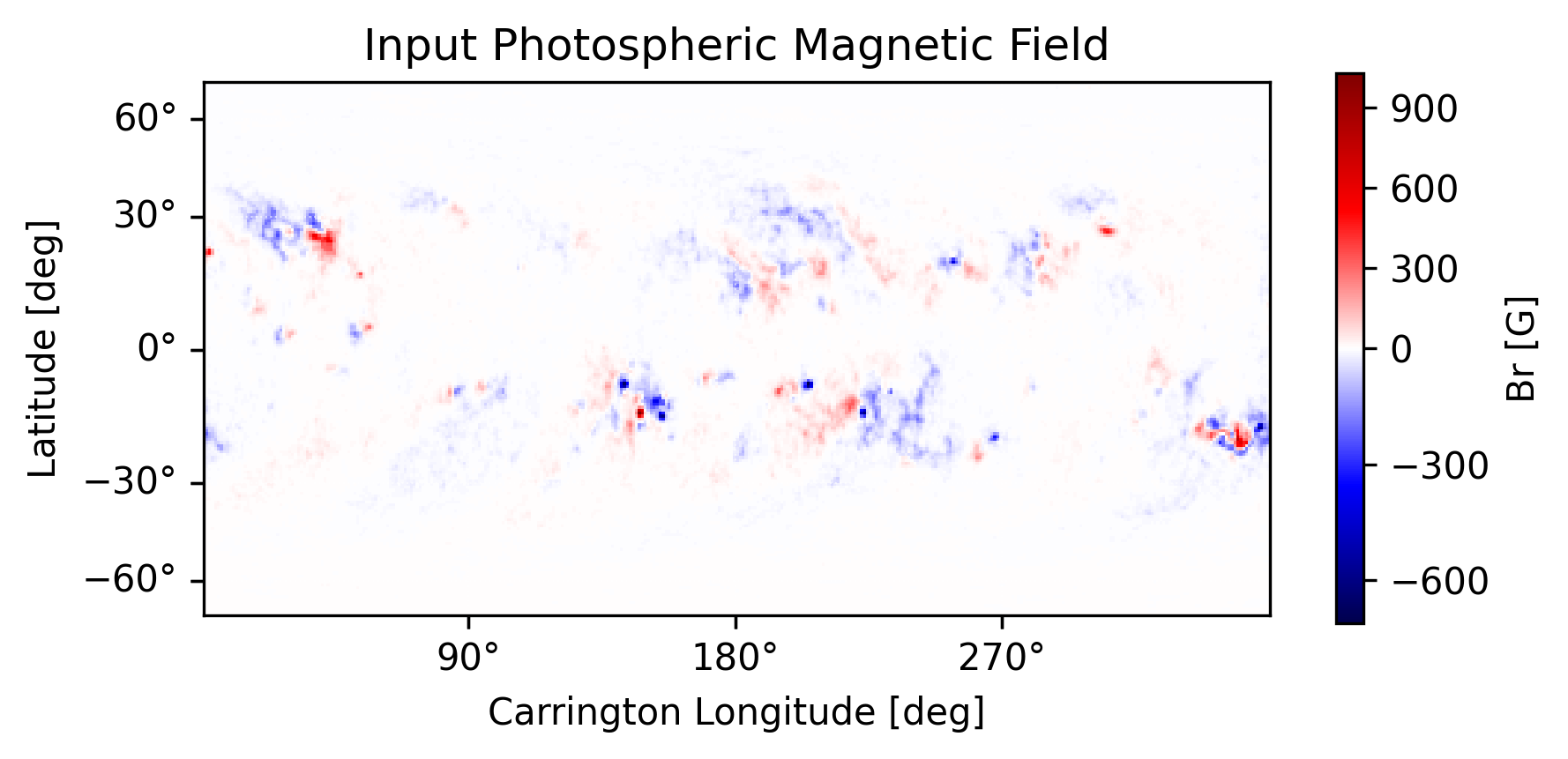}
    \caption{GONG radial magnetic field CR 2284 synoptic map. The map is used as a lower boundary condition for the \texttt{pfsspy}. }
    \label{input-field-br.png}
\end{figure}

\begin{figure}
    \centering
    \includegraphics[width=1\linewidth]{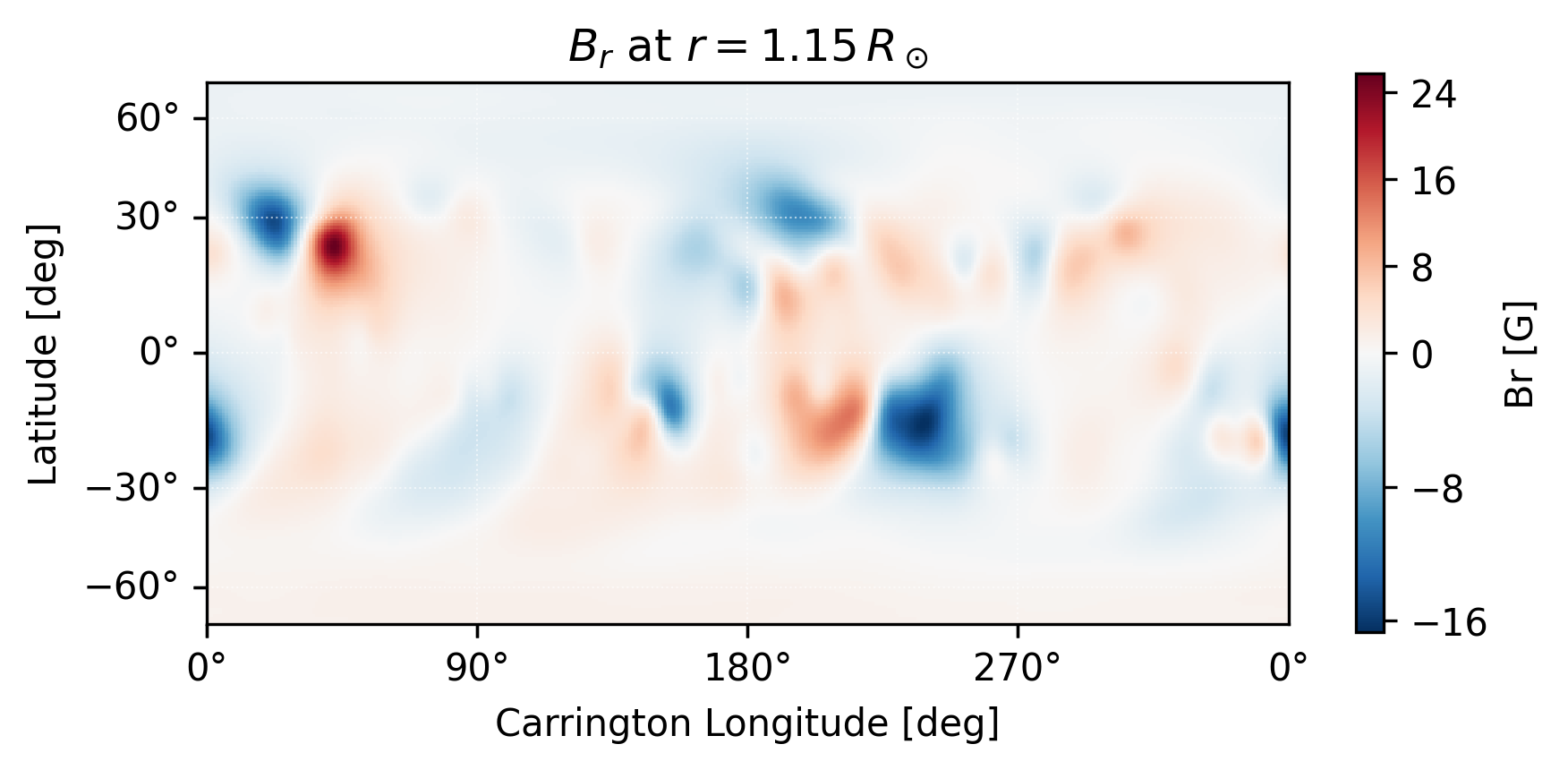}
    \caption{Contour plot of the \texttt{pfsspy} output radial magnetic field at some height reference equal to r=1.15 R$_{\odot}$.}
    \label{PFSS-at-1.15.png}
\end{figure}

We used the The Global Oscillation Network Group (GONG) Carrington rotation 2284 radial magnetic field to compute the initial PFSS field from the photosphere up to a source surface at 2.5 R$_\odot$. The PFSS was computed using the \texttt{python} package \texttt{pfsspy} \citep{Stansby2020-PFSSPY}, with a resolution of $360 \times 180 \times 37$ along $\phi$, $\theta$ and $r$ coordinates, respectively. The resolution can also be changed depending on the goals. A contour plot of the input (photospheric) radial component of the magnetic field from the \texttt{pfsspy} is plotted in Fig. \ref{input-field-br.png}. In Fig. \ref{PFSS-at-1.15.png}, we plot the radial component of the magnetic field at 1.15 R$_\odot$ which we will use as a reference for a more qualitative view of the impacts of the optimization on the magnetic field. We choose $r=1.15$ R$_\odot$ because it is close enough to the photosphere to capture the impacts of the majority of the loops included, since they have different lengths, ranging from $\approx$ 0.15 to 1.60 R$_\odot$. The computed PFSS is the input field to the method explained in section \ref{multi-constrained-model-method-section}.\par


\begin{figure}
    \centering
    \includegraphics[width=1\linewidth]{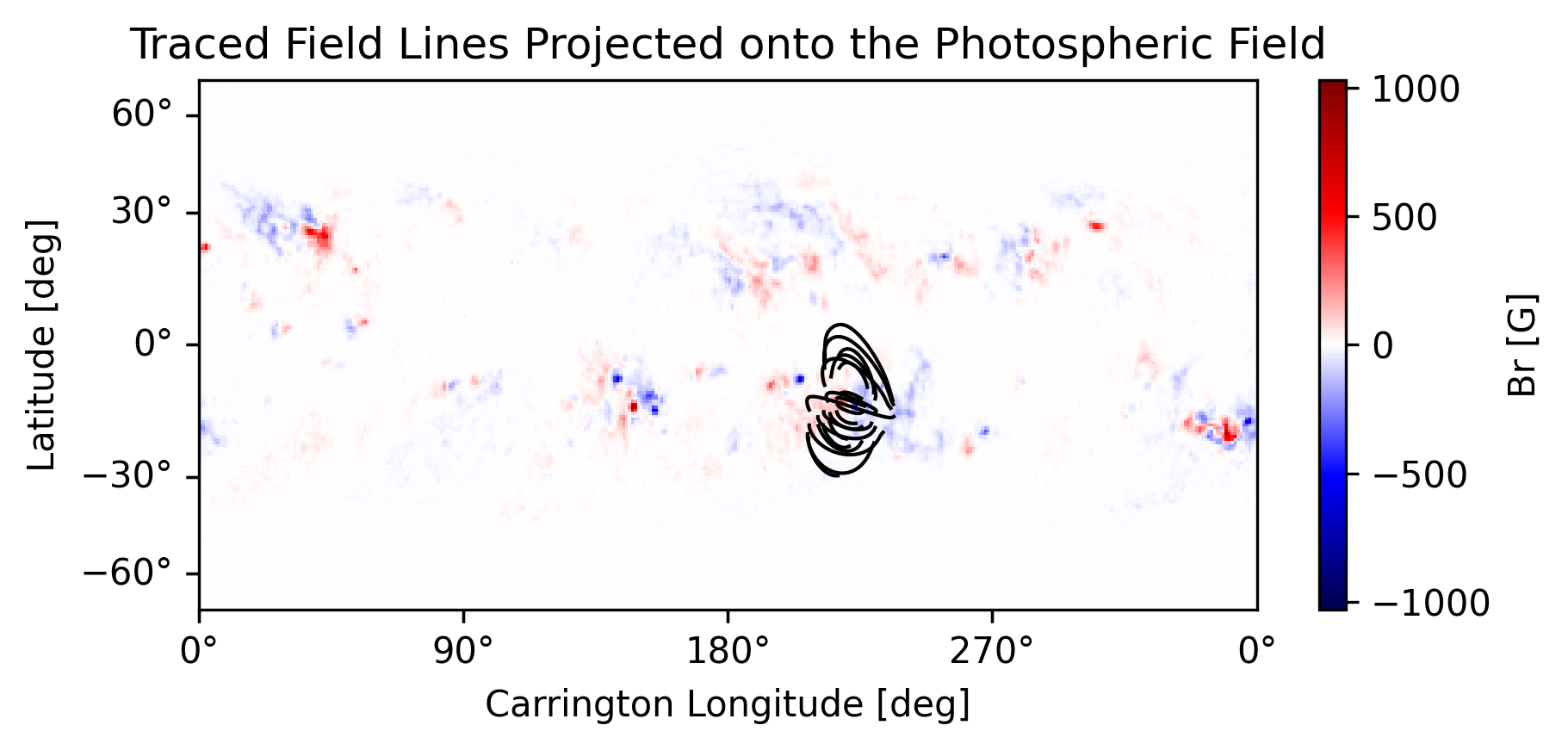}
    \caption{Projection of the selected PFSS loops traced using the \texttt{pfsspy} output magnetic field onto the CR 2284 map radial photospheric magnetic field.}
    \label{projected-loops-photosphere.png}
\end{figure}

\begin{figure}
    \centering
    \includegraphics[width=1\linewidth]{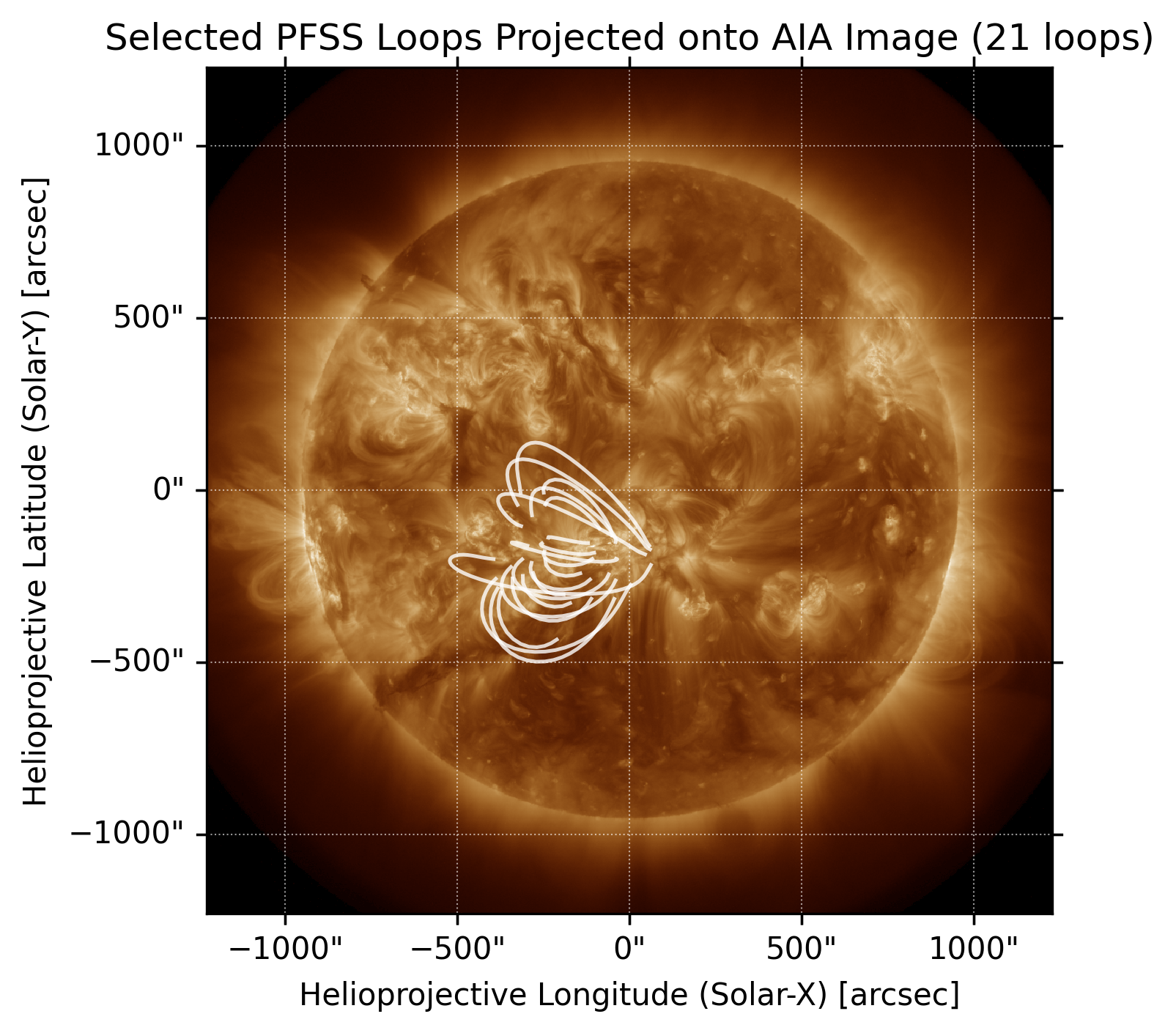}
    \caption{Projection of the selected (non-modified) traced loops on SDO/AIA 193 $\AA$ image taken on 2024 May 16 at 13:00 UT.}
    \label{aia-pfss-original-selected-loops}
\end{figure}

\begin{figure}
    \centering
    \includegraphics[width=1\linewidth]{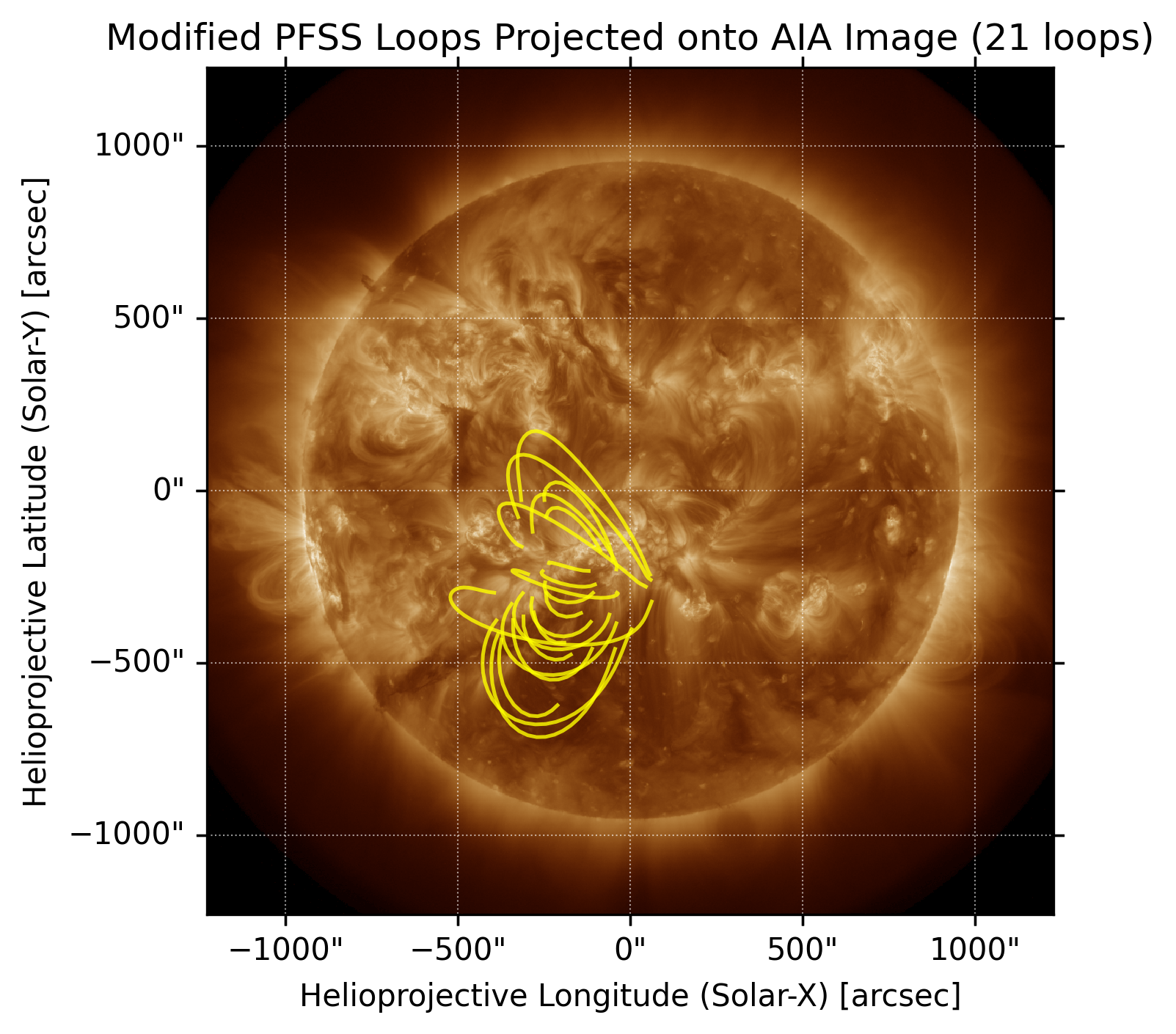}
    \caption{Projection of the modified loops on the SDO/AIA 193 $\AA$ EUV image taken on 2024 May 16 at 13:00 UT. The modified loops, shown in yellow, correspond to those in Fig. \ref{aia-pfss-original-selected-loops} whose heights were artificially increased by a factor of 1.4.}
    \label{aia-modified-loops-projection}
\end{figure}

Based on the output from the \texttt{pfsspy}, we trace some loops using the \texttt{FortranTracer}, which is included in the \texttt{pfsspy} package. We modify the loops by increasing their height by a factor of 1.4. We then select some of the traced loops by excluding those associated with larger angles, closer to 80/90º, and selecting the ones described by at least six points. The selected set of 21 loops is projected onto the photosphere (Fig. \ref{projected-loops-photosphere.png}) and onto the SDO/AIA 193 $\AA$ image taken on 2024 May 16 at 13:00 UT, showing the EUV emission of the target region (Fig. \ref{aia-pfss-original-selected-loops}). The corresponding modified loops are shown on the same AIA image (Fig. \ref{aia-modified-loops-projection}). In Figs. \ref{aia-pfss-original-selected-loops} and \ref{aia-modified-loops-projection}, the loops are projected as they would appear from the SDO spacecraft viewpoint, according to the \texttt{python} routine in \citet{Stansby2020-PFSSPY}. \par

\section{Results and Discussion}\label{results-section}

The modification of the original loops has an effect on the force-freeness of the solution; that is, one can achieve better alignment between the modified loops and the magnetic field at the expense of a worse force-free and/or divergence degree. As changing the original loops leads to a trade-off between both the divergence and force-free constraint and the $L_2$ term, we adapted the model to be able to "switch on and off" the different constraints ($L_2$ and $L_3$). This allows the code to evaluate for either an $L_t=L_1$ or an $L_t=L_1+\xi_3L_3$ minimum, depending on whether the force-free constraint is included, by taking advantage of the flexibility of the model in including and excluding constraints. When $L_2$ is "switched off", no interpolations back to the grid are performed, $F_2$ are always zero along the loops, and $L_2$ are simply accessed at every 10 iterations but not used to define the success or rejection of an iteration. This decreases the required computation time due to unnecessary interpolation, which is the most time-consuming step. \par

The details regarding the two analyzed configurations can be found in Table \ref{tab:cases}. In Setup I, we set $\mu_3=0$, meaning $L_3$ is assessed but not used to evaluate iteration success. Initially $L_2$ is active, the number of constraints is $N=2$, and an iteration is successful when $L_1(t+\Delta t)+L_2(t+\Delta t)<L_1(t)+L_2(t)$. After a minimum is reached, with no further functional decrease, $L_2$ is switched off, and the iteration continues, whose success is now based solely on whether $L_1(t+\Delta t)<L_1(t)$. For Setup II we use the same procedure, but contrary to Setup I, $L_3$ is always evaluated and used to define an iteration success or rejection. Initially $L_2$ constraint is active, the total number of constraints is $N=3$, and the full functional $L_t = L_1 + L_2 + \xi_3 L_3$ is minimized. After $L_2$ is switched off, the code again proceeds until a new $L_t = L_1 + \xi_3 L_3$ minimum is obtained. \par

\begin{table*}[htbp]
\centering
\caption{Summary of the two optimization setups used in this study. Each setup includes two phases, depending on whether the tangential constraint term $L_2$ is switched on or off. The terms $L_1, L_2$, and $L_3$ represent the solenoidal, tangential, and force-free constraints, respectively. "Yes" indicates that the corresponding constraint is included in the minimization, and "Assessed only" that it is evaluated but not used in determining iteration success. The last column specifies the criterion defining a successful iteration during each phase.}
\resizebox{\textwidth}{!}{
\begin{tabular}{lccccl}
\hline\hline
\textbf{Setup/case} & \textbf{$L_2$ Contribution} & $L_1$ & $L_2$ & $L_3$ & \textbf{Iteration success criterion} \\
\hline
\multirow{2}{*}{I} 
  & $L_2$ switched on (until reaching a minimum)  & Yes & Yes & Assessed only & $L_1(t+d t) + L_2(t+d t) < L_1(t) + L_2(t)$ \\[2pt]
  & $L_2$ switched off (afterward)  & Yes & Assessed only  & Assessed only & $L_1(t+d t) < L_1(t)$ \\[2pt]
\hline
\multirow{2}{*}{II}
  & $L_2$ switched on (until reaching a minimum)  & Yes & Yes & Yes & $L_1(t+d t) + L_2(t+d t)+\xi_3L_3(t+d t) < L_1(t) + L_2(t) +\xi_3L_3(t)$ \\[2pt]
  & $L_2$ switched off (afterward) & Yes & Assessed only  & Yes & $L_1(t+d t) +\xi_3L_3(t+d t) < L_1(t) +\xi_3L_3(t)$\\[2pt]
\hline
\end{tabular}}
\label{tab:cases}
\end{table*}

For both setups, we plot the constraints $L_n$ as a function of the number of iterations normalized to the initial value at iteration $t=0$ for better comparison (Figs. \ref{Ln-vs-noa-only-L1} and \ref{ln-vs-noa-15-iter.png} for cases I and II). The root mean square of the final angles between the tangents to the 21 included loops can be found in Fig. \ref{final-angles-rms}. Based on the output optimization perturbed magnetic field, we trace the field lines at the same footpoints of Figs. \ref{projected-loops-photosphere.png} and \ref{aia-pfss-original-selected-loops}, and project them again onto the SDO/AIA image (Figs. \ref{perturbed-pfss-only-l1-AIA.png} and \ref{perturbed-pfss-on-aia-final-1.15.png}). The Fig. \ref{1.15-perturbed-1.15-final.png} is the contour plot of the output magnetic field at 1.15 R$_\odot$. The plots from the end of the $L_2$ switch-off period are shown in the following two subsections.

\subsection{ Setup I: $N=2$ Followed by a $L_2$ Switch Off}\label{section-case-n=2}

\begin{figure}
    \centering
    \includegraphics[width=1\linewidth]{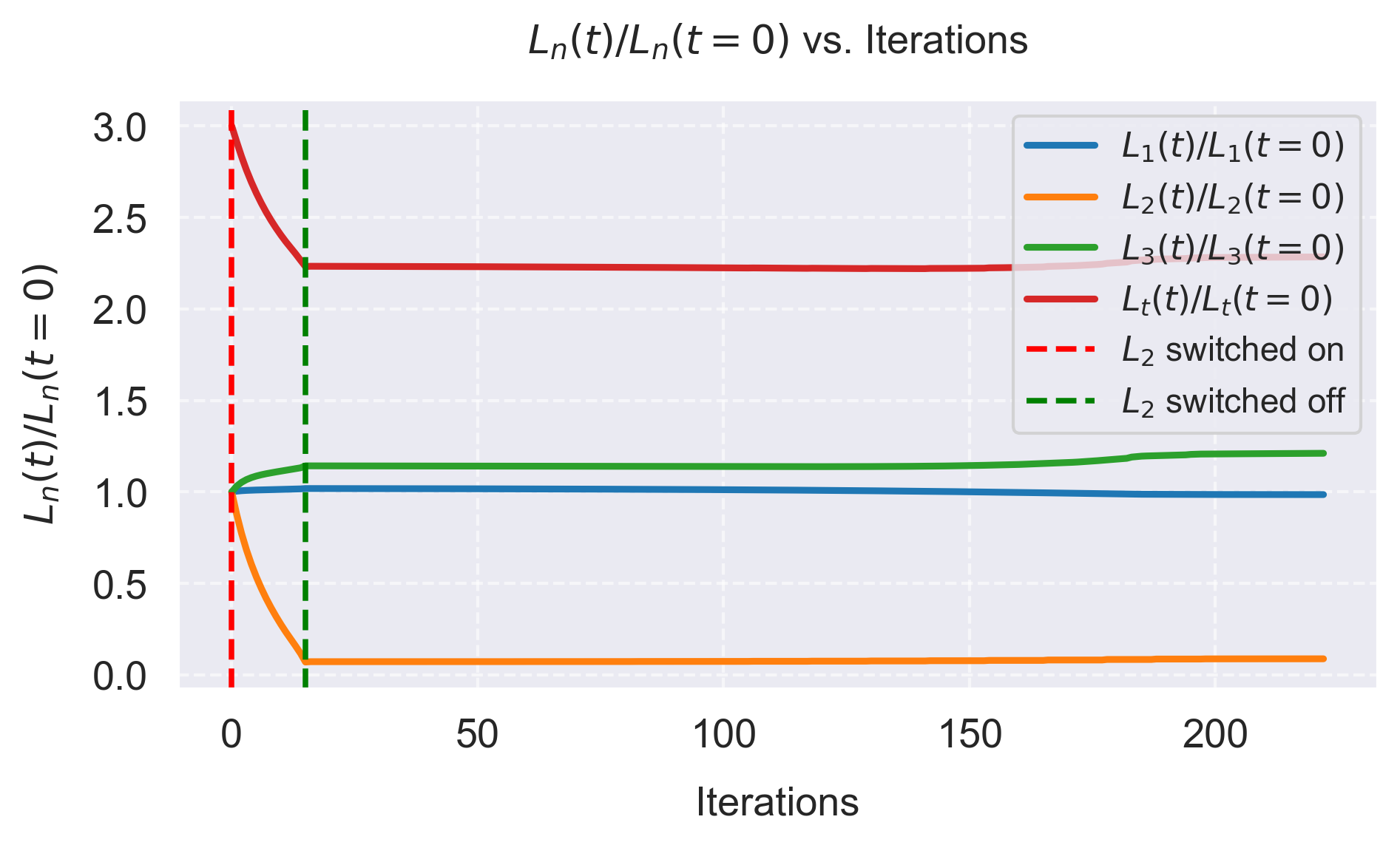}
    \caption{Normalized functional terms $L_n / L_n(t=0)$ as a function of the number of iterations for case I. The vertical green dashed line marks the point at which the $L_2$ constraint is switched off.}
    \label{Ln-vs-noa-only-L1}
\end{figure}
For case I, the initial $\mu_1$ was set to $10^{-4}$ and $\mu_2=0.5$, with $\xi_1=\xi_2=1$. A higher initial value of $\mu_2$ is used because the threshold at which $L_2$ starts decreasing is higher than that of both $L_1$ and $L_3$; this choice improves the computational efficiency. The code required approximately 5 minutes and around 260 iterations to converge to a solution. The code ran on an Apple M3 processor. In Fig. \ref{Ln-vs-noa-only-L1}, an initial decrease in $L_2$ is observed, which means that the method successfully reduces the angle between the interpolated magnetic field and the tangents to the included loops. Such a trend is accompanied by an increase of both $L_1$ and $L_3$. This is a result of the inclusion of loops that disturb the original field, creating field divergence (see also the initial disturbance in Figs. \ref{local-l1-merged} and \ref{local-l3-merged}). A better tangency of the magnetic field to the included loops is achieved by means of a worsening of the divergence and force-freeness degree. During the $L_2$-off phase, the algorithm simply evaluates for an $L_1$ minimum, reached after around 220 successful iterations. The curve attenuation associated with the minimum can be seen more clearly in Fig. \ref{local-l1-merged} from the appendix \ref{local-plots}. After the $L_2$ switch-off, the divergence decreases monotonically. $L_3$ remains at the levels of its necessary initial rise for accommodating the loops, except closer to the minimum, after which the divergence and force-freeness tend to become a trade-off between each other.

\begin{figure}[H]
    \centering
    \includegraphics[width=1\linewidth]{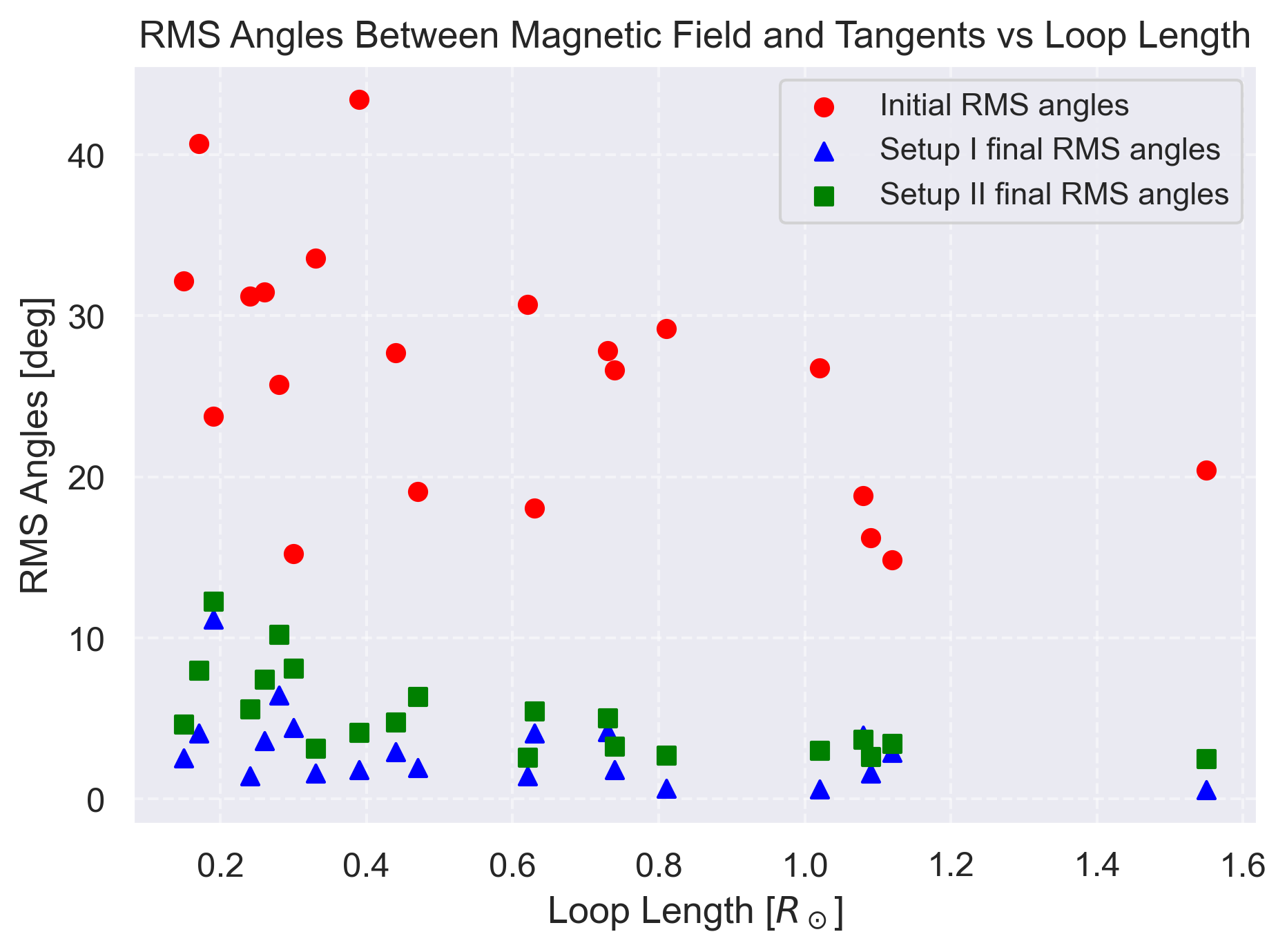}
    \caption{Root mean square (RMS) of the initial (red dots) and final angles (green squares and blue triangles) between the interpolated magnetic field and the tangents to the loops, shown as a function of the geometric length of each loop (in units of R$_\odot$), for the two optimization setups and the 21 loops used. Smaller RMS values correspond to a closer alignment between the magnetic field and the coronal loop geometry.}
    \label{final-angles-rms}
\end{figure}

Based on Figs. \ref{Ln-vs-noa-only-L1} and \ref{final-angles-rms}, the method yields magnetic field solutions that allow for non-force-free field configurations in the vicinity of the included loops, while maintaining low divergence levels across the computational domain and good alignment of the final magnetic field. After switching off the $L_2$ term, the magnetic field alignment to the loops slightly worsens. The $L_2$ switched-off phase compensates for the increase in divergence introduced by the loop inclusion (see Fig. \ref{local-l1-merged}). \par

\begin{figure}
    \centering
    \includegraphics[width=1\linewidth]{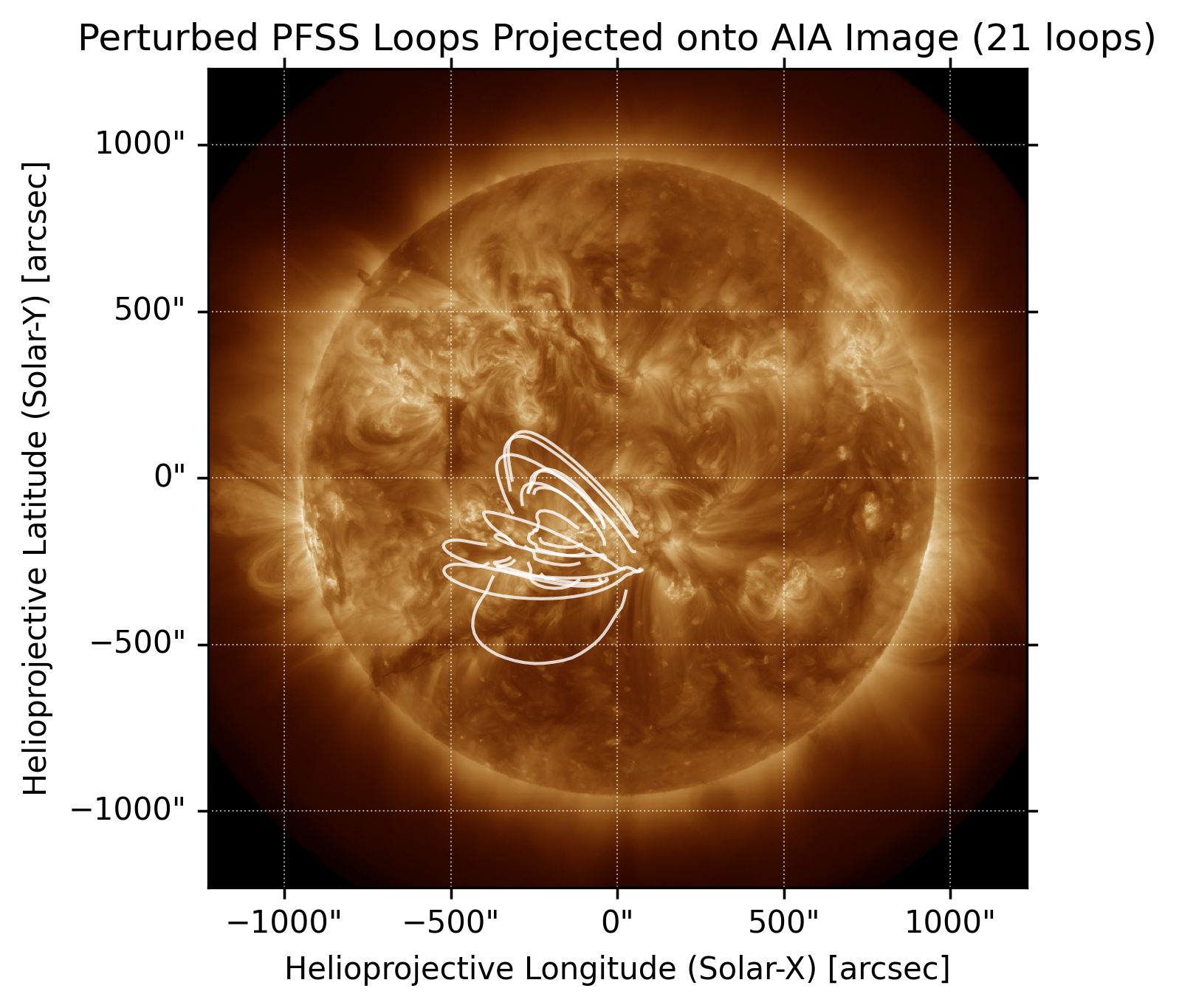}
    \caption{Projection of the magnetic field lines traced from the original footpoints using the optimized magnetic field (setup I) onto an SDO/AIA $193$ $\AA$ EUV image taken on 2024 May 16 at 13:00 UT. The selected footpoints correspond to the same set used in Fig. \ref{aia-pfss-original-selected-loops}.}
    \label{perturbed-pfss-only-l1-AIA.png}
\end{figure}

\begin{figure}
    \centering
    \includegraphics[width=1\linewidth]{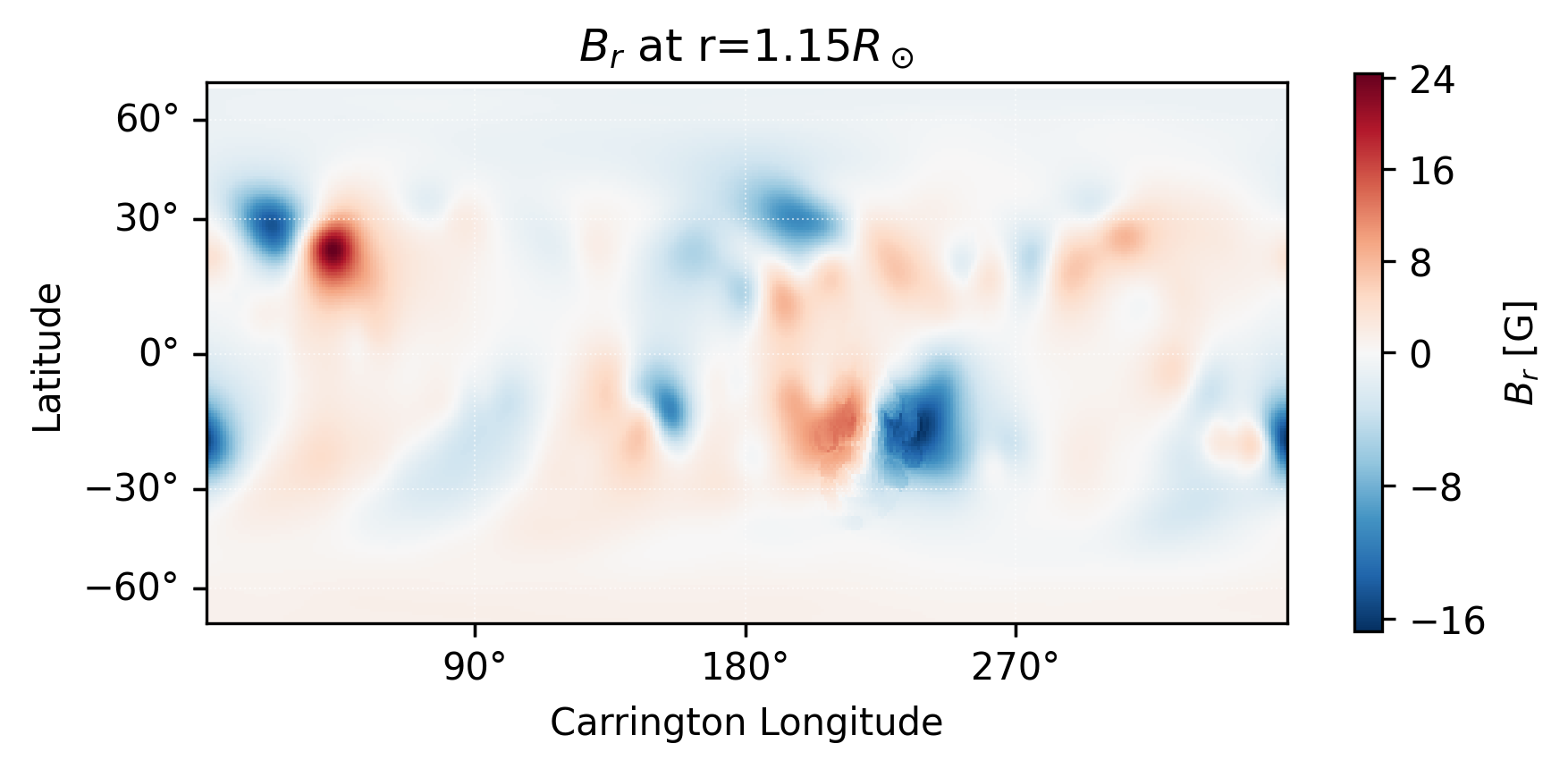}
    \caption{Contour map of the optimized radial magnetic field component, $B_r$, evaluated at the reference height $r = 1.15$R$_\odot$ for case II. }
    \label{1.15-perturbed-1.15-final.png}
\end{figure}

In the radial component of the magnetic field contour plot at r$=$1.15 R$_\odot$ only changes of the magnetic field occur close to the regions where the loops are included. The contour plot is identical to that from case II represented in Fig. \ref{1.15-perturbed-1.15-final.png}, where disturbances can be seen in the region where the loops are present. Tracing field lines at the original footpoints, plotted in Fig. \ref{perturbed-pfss-only-l1-AIA.png}, do change, comparatively to the original, undisturbed field and field lines from Fig. \ref{aia-pfss-original-selected-loops}. The field lines projected onto the AIA image are significantly different from the original ones as seen in Fig. \ref{aia-pfss-original-selected-loops}. Even though some of the loops included are distinct from the original ones, they still influence the nearby regions where the "non-modified" loops are present.

\subsection{ Setup II: $N=3$ Followed by a $L_2$ Switch Off }

Since the $\mu_n$ thresholds change with the inclusion of the new constraint, the initial $\mu_1$ and $\mu_3$ are set to $10^{-5}$ while $\mu_2=0.5$, and $\xi_1/\xi_3=2$, following the procedures described in \cite{Wiegelmann2004}. A computation time of $\approx$ 3 minutes and 40 seconds on an Apple M3 was required for 140 iterations, after which $L_t$ becomes stationary.\par

An initial rise of $L_1$ and $L_3$ in the Fig. \ref{ln-vs-noa-15-iter.png} is noticeable (see Fig. \ref{local-l1-merged} in appendix \ref{local-plots}). This means the magnetic field requires passing through a series of non-force-free states with higher divergence before improving its force-freeness and divergence degree for a better tangency of the magnetic field to the included loops, a trend also seen in setup I. In this phase, the loop influence dominates locally over the other constraint adjustments and causes the initial disturbance, the rise of both $L_1$ and $L_3$. If the loops are included at $t>0$, the same rising patterns in $L_1$ and $L_3$ constraints are obtained. In such a case, only a small difference in the $L_1$ and more noticeably in $L_3$ values of the final state (after $L_2$ switch off) is obtained compared to the results in Figs. \ref{Ln-vs-noa-only-L1} and \ref{ln-vs-noa-15-iter.png}. Unless the number of loops included is not significantly increased, no significant changes in the final $L_1$ (and $L_3$) are expected. Since $L_1$ and $L_3$, are global constraints which are fulfilled by means of local adjustments, the success of each iteration becomes dependent on the decrease of both in a large fraction of the computational domain when $L_2$ is switched off. By including the loops at the beginning of the optimization, the magnetic field is influenced by and adjusted consistently to accommodate the loops towards the minimum. \par

\begin{figure}
    \centering
    \includegraphics[width=1\linewidth]{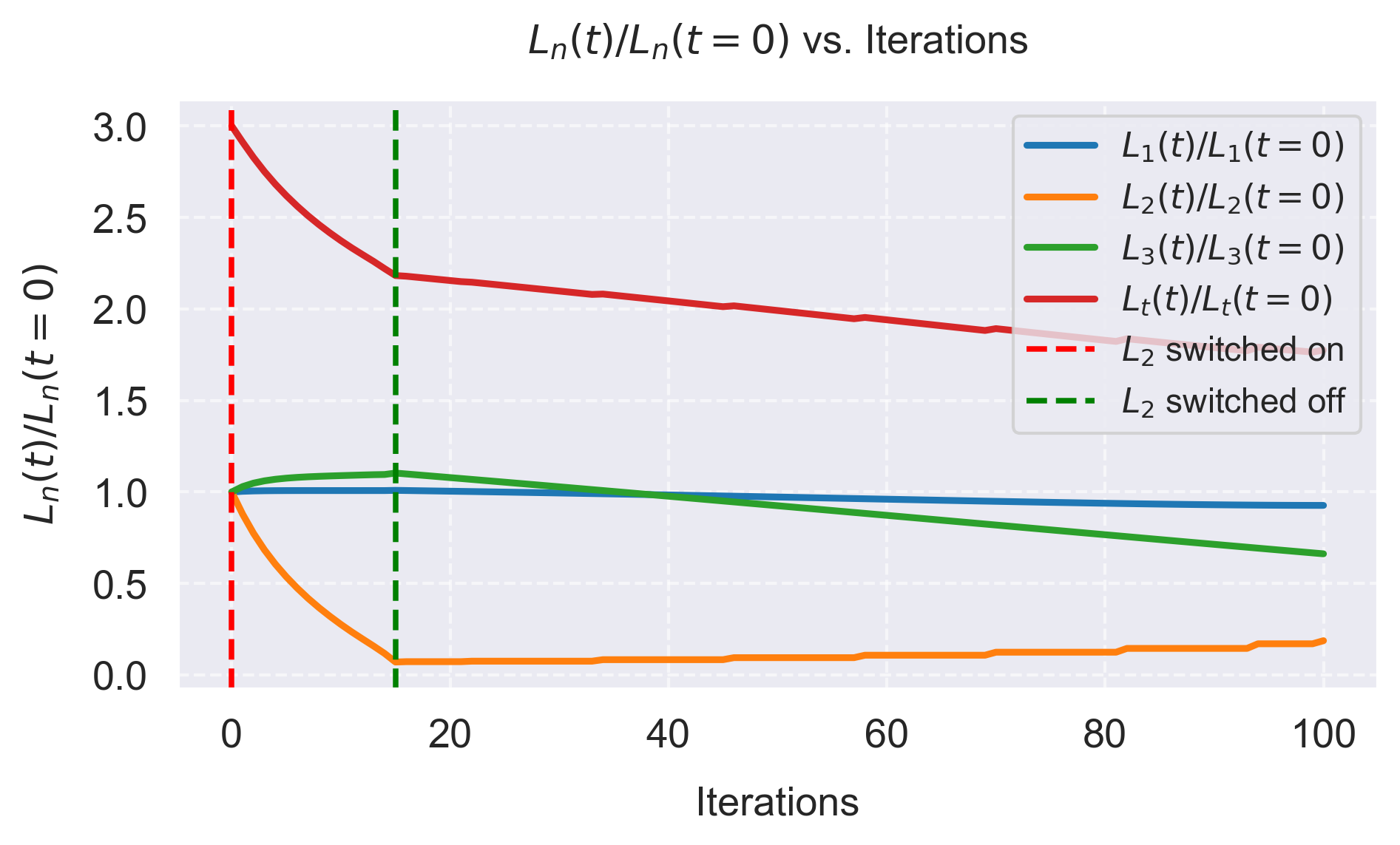}
    \caption{Normalized functional terms $L_n / L_n(t=0)$ as a function of the number of iterations for case II. The vertical green dashed line marks the point at which the $\boldsymbol{L}_{\mathbf{2}}$ constraint is switched off.}
    \label{ln-vs-noa-15-iter.png}
\end{figure}

\begin{figure}
    \centering
    \includegraphics[width=1\linewidth]{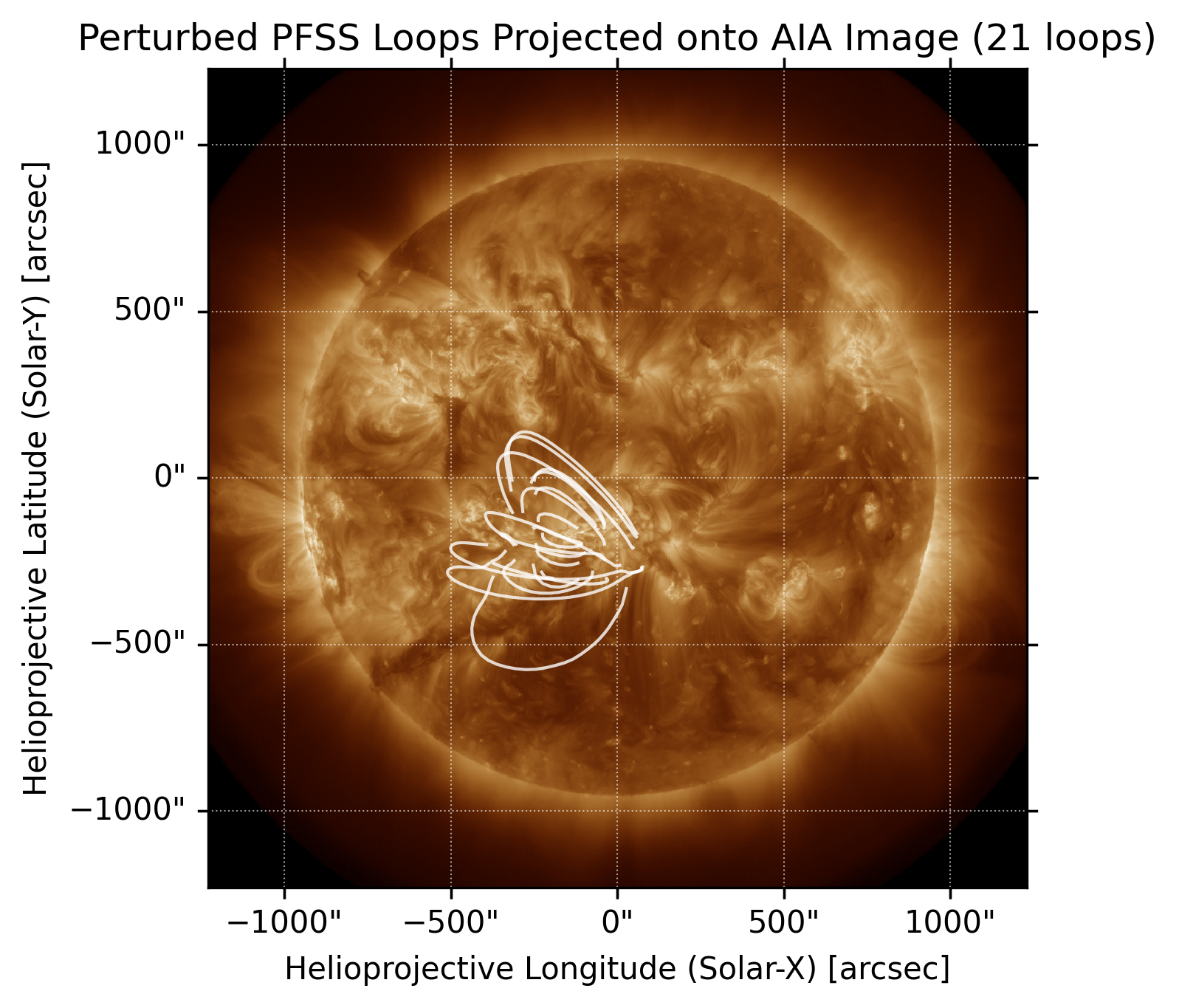}
    \caption{Projection of the magnetic field lines traced from the original footpoints using the optimized magnetic field (setup II) onto an SDO/AIA $193$ $\AA$ EUV image taken on 2024 May 16 at 13:00 UT. The selected footpoints correspond to the same set used in Fig. \ref{aia-pfss-original-selected-loops}.}
    \label{perturbed-pfss-on-aia-final-1.15.png}
\end{figure}

In the $L_2$-switched-off phase, the divergence in regions close to the loops (approximate cylindrical volumes enclosing the loops, with a radius of $\approx 0.1$ R$_\odot$) also decreases (see Fig. \ref{local-l1-merged} from appendix \ref{local-plots}), but only down to the level before the loop inclusion. The divergence decrease is more pronounced than in Setup I, both local and globally (see also Fig. \ref{local-l1-merged} from appendix \ref{local-plots}). Therefore, the $L_3$ force-freeness constraint also helps decrease the divergence of the magnetic field. This is likely related to the solution being closer to the initial undisturbed state and magnetic field stress removal at the expense of the field-tangent alignment. Similarly to the case I, an increase in the final alignment between the magnetic field and the included coronal loops is still obtained, as seen in Fig. \ref{final-angles-rms}, although slightly worse than that of case I because the force-freeness of the field increases. The loop-field tangency may evolve differently depending on the loops. Shorter loops relative to the grid size, which are sampled by fewer points, tend to be more challenging for the optimization. This likely explains the larger final RMS angles for shorter loop lengths, as seen in Fig. \ref{final-angles-rms}. Adding more loops may, therefore, have an effect on the loop-field tangency and on the $L_2$ decrease rate. This could also happen in cases when there is an overlap of loop influence zones, or when the initial loop-magnetic field alignment is already good. The field lines projection onto AIA shown in Fig. \ref{perturbed-pfss-on-aia-final-1.15.png} shows differences, especially for shorter loops when compared to Fig. \ref{perturbed-pfss-only-l1-AIA.png} from the setup I. \par

As also seen in Fig. \ref{ln-vs-noa-15-iter.png}, the inclusion of a force-freeness constraint helps guiding the field toward a higher force-freeness, lower divergence state, while still providing enough "freedom" for the field to accommodate loop inclusion, departing from the potential state in regions close to the loops. The $L_3$ term reduces the currents misaligned with the magnetic field, increases divergence-freeness, and makes the process more computationally efficient. The improvement of the solenoidal and force-free constraints comes at the expense of a reduced alignment between the magnetic field and the loop tangents compared to setup I. The loop-field alignment does not revert to its initial state, so the final configuration remains more consistent with the observations and significantly different from the initial one in the vicinity of the loops. Although $L_1, L_3$, and $L_2$ exhibit a trade-off, this trade-off becomes weaker in the subsequent $L_2$-off phase than when the $L_2$ constraint is active.

\subsection{Finding the Optimal $\xi_2$ }

\begin{figure}
    \centering
    \includegraphics[width=1\linewidth]{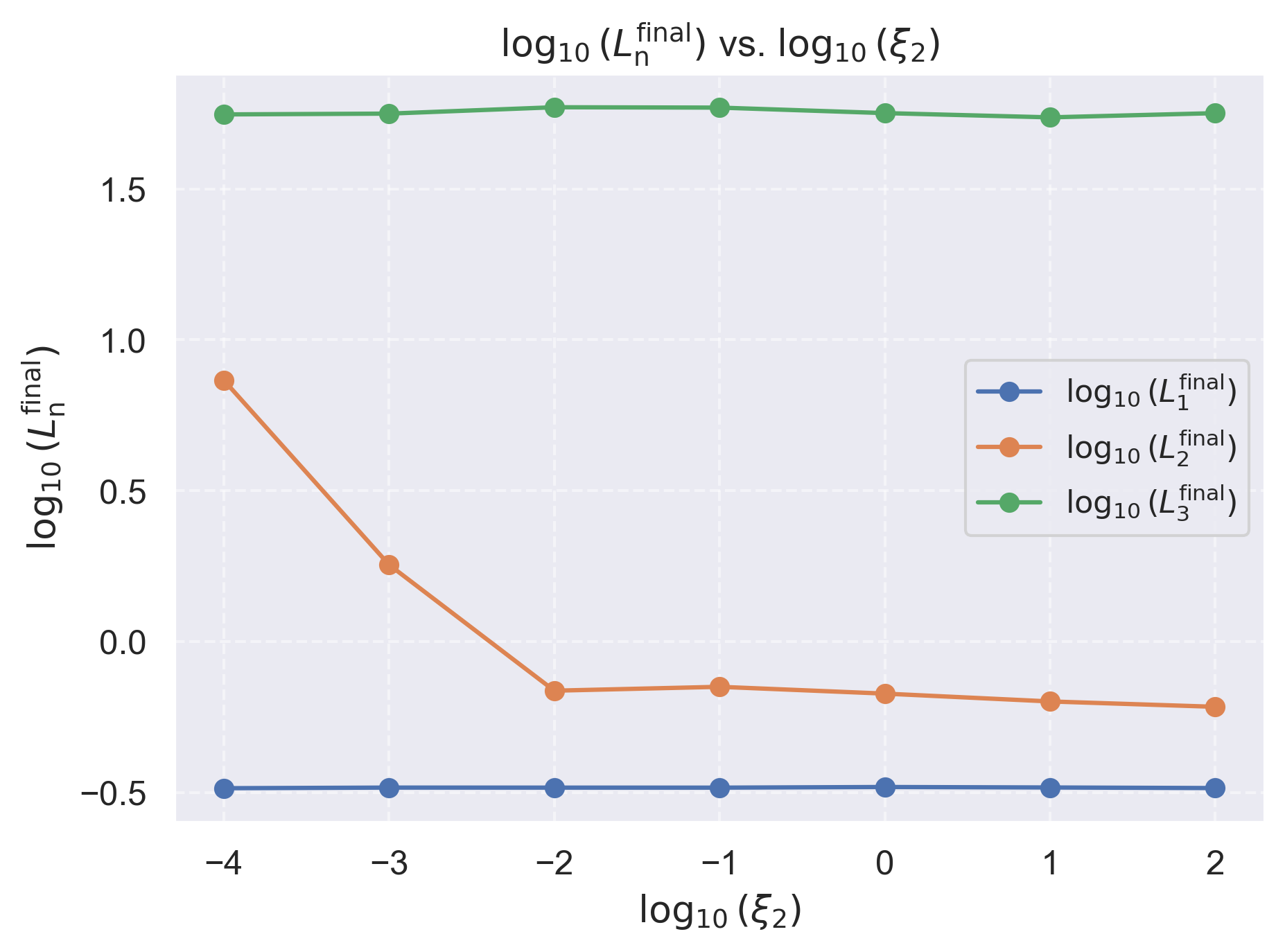}
    \caption{Logarithmic plot of final functionals $L_1, L_2$, and $L_3$ as functions of the regularization parameter $\xi_2$, for Setup I after the $L_2$ constraint has been switched off (i.e, at the achieved minimum). }
    \label{qsi-vs-/ln-vs-qsi2-final-only-l1}
\end{figure}

\begin{figure}
    \centering
    \includegraphics[width=1\linewidth]{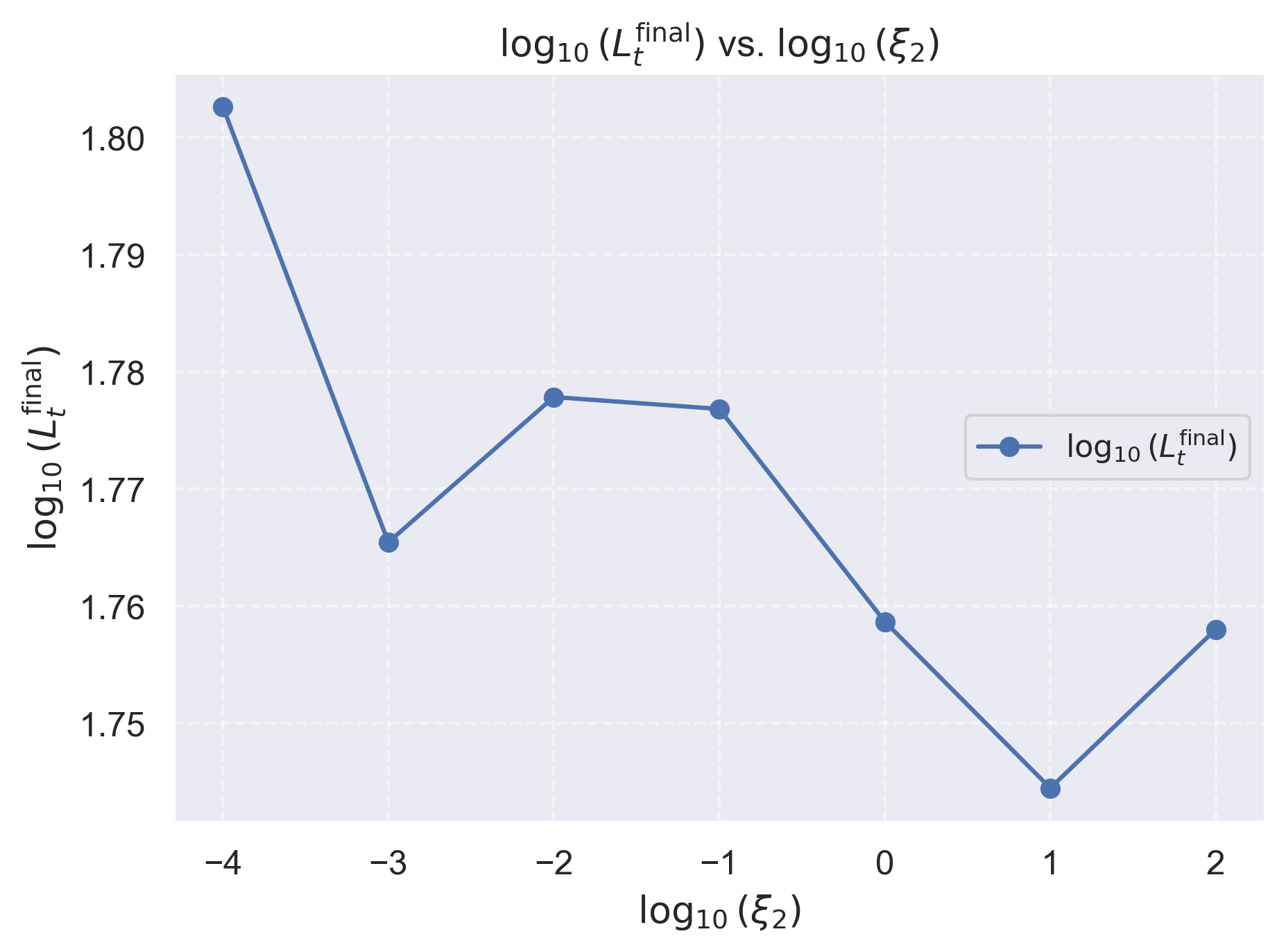}
    \caption{Logarithmic plot of the final $L_t$ (the sum of the three functional terms in figure \ref{qsi-vs-/ln-vs-qsi2-final-only-l1}) as functions of the regularization parameter $\xi_2$, for Setup I after the $L_2$ constraint has been switched off (i.e, at the achieved minimum).}
    \label{total_lt_setupIvsqsi}
\end{figure}

To find the optimal $\xi_2$ and assess the dependence of the model on the choice of $\xi_2$, we plotted the $L_n$ terms as a function of $\xi_2$, in Fig. \ref{qsi-vs-/ln-vs-qsi2-final-only-l1}. $L_1$ and $L_2$ are insensitive to $\xi_2$ as long as $\xi_2 \gtrsim 10^{-2}$. This occurs likely because the regions influenced by the loops are much smaller than the total computational volume. For $\xi_2<10^{-2}$, the loop-field tangency can be controlled by choosing how large $\xi_2$ is. Smaller $\xi_2$ values require a larger number of iterations and computation times. Based on the final values of the final three-term functional in Fig. \ref{total_lt_setupIvsqsi}, we conclude the optimal $\xi_2$ value is around $10$ for Setup I. For setup II, very similar trends are found, but the optimal value is larger: $\xi_2 \approx 10^2$. For $\xi_2$ values below unity in setup II, the cumulative influence of the $L_1$ and $L_3$ constraints competes with that of $L_2$, dominating over the smaller $\xi_2 F_2$ adjustments. Therefore, the optimal regularization parameters depend on both the number of active constraints and the relative rate of change among $L_1$, $L_2$, and $L_3$, if $\xi_2$ is not large enough.

\section{Conclusions }\label{conclusions}

We show a method to include 3D coronal loop shape information into the PFSS. By using a second-order finite difference algorithm and imposing a potential lower boundary derived from photospheric observations, it is possible to obtain a magnetic field solution with lower divergence but also non-force-freeness levels (depending on the applied constraints) while tangent to the included loops. The method simultaneously incorporates, incrementally, the 3D information related to the shape of coronal loops while also being constrained at the photospheric level. The inclusion of a divergence constraint successfully mitigates the disturbances introduced by loop inclusion. Computational efficiency is preserved even when a large number of loops are added. Analogously, the inclusion of a force-free constraint controls the Lorentz forces created by loop inclusion, increasing the force-freeness degree and also helps find more divergence-free solutions.\par

This work paves the way for a faster, more physically consistent approach to obtaining the global coronal magnetic field. The code is designed to incorporate real, stereoscopically reconstructed loops as new observations become available, and thus also serves as a key testbed for future developments. The inclusion of loops traced by other extrapolation models is also a possibility. By selectively including or excluding specific constraints, the method allows the solution to be tuned in terms of its loop-field tangency and force-freeness degrees. With explicit control over the tangency of field lines to the observed coronal loops, the model could enable the control and quantification of the amount of free magnetic energy introduced by the included loop or system of loops. This, in turn, would make it possible to analyse the resulting free-energy distribution and rearrangement. Moreover, chromospheric or other observational constraints could be analogously integrated by similar procedures, depending on the resolution. Future work should focus on performing detailed comparisons with other extrapolation techniques. The dependence of the method on the magnetic-field complexity, namely on the differences between solar-maximum and solar-minimum cases in obtaining solutions that satisfy fixed boundary conditions, should also be further investigated. The inclusion of real coronal loops into the new PFSS-optimization framework, along with an associated magnetic energy study, is currently underway.

\section*{Acknowledgements}
C.A. and R.G. thanks to Fundação para a Ciência e a Tecnologia (FCT) through the research grants UUID/04434/2025. 
I.C. acknowledges the support of the Coronagraphic German and US Solar Probe Plus Survey (CGAUSS) project for WISPR by the German Aerospace Center (DLR) under grant 50OL2301.
We thank P. Rasori (WorldTalents company) for funding \& mentorship activities."

\bibliographystyle{apalike} 
\bibliography{biblio.bib} 

\begin{appendix}

\section{Effect of Switching $L_2$ On/Off on the Local $L_1$ and $L_3$}\label{local-plots}

\begin{figure}[H]
    \centering
    \includegraphics[width=1\linewidth]{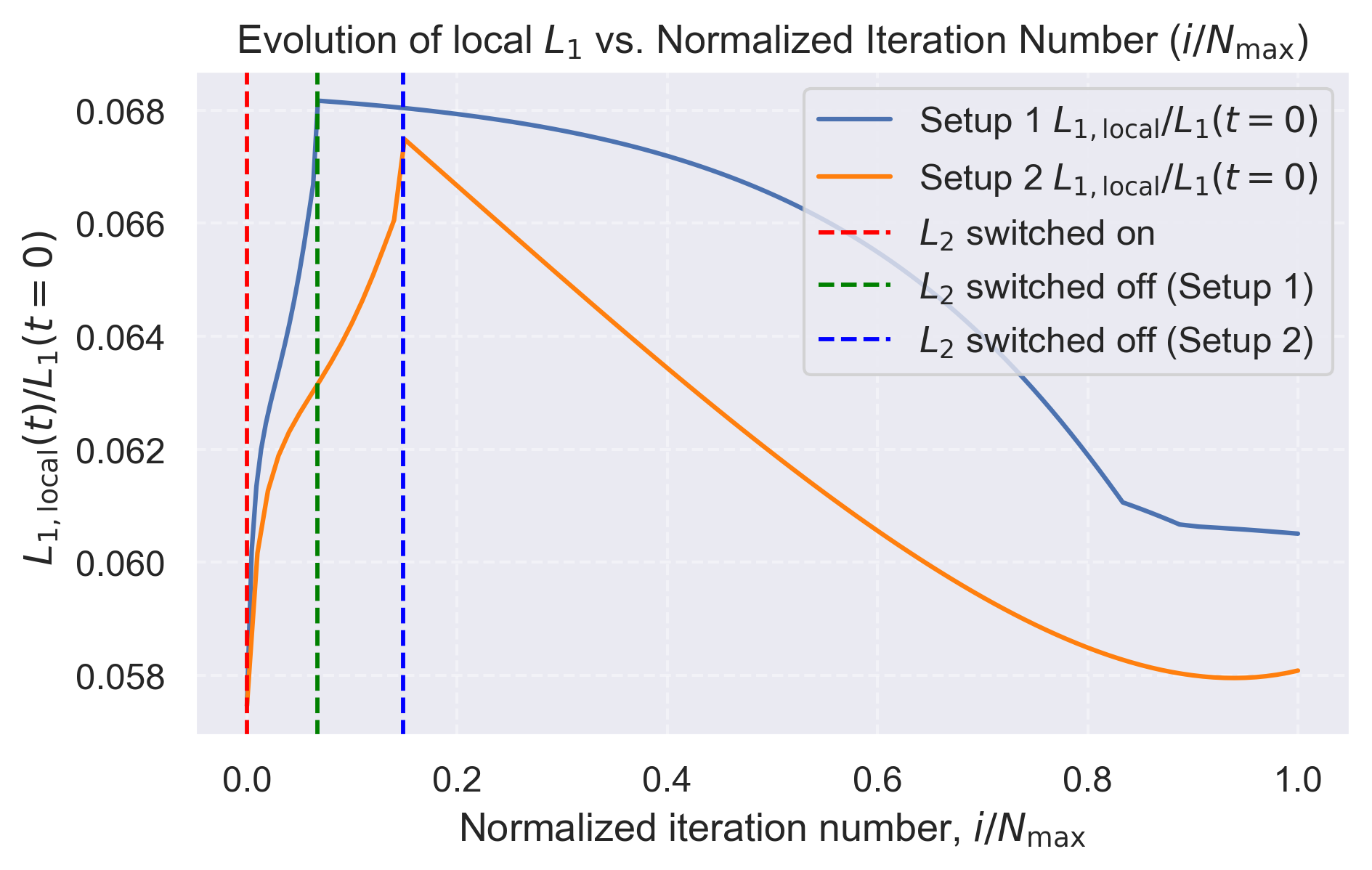}
    \caption{Evolution of the local metric $L_{1, \text{local}}$, normalized to its initial value, as a function of the normalized iteration number $i / N_{\text{max}}$ for Setup 1 (blue curve) and Setup 2 (orange curve). The vertical red dashed line indicates the iteration at which $L_2$ is switched on, while the green and blue dashed lines mark the iterations at which $L_2$ is switched off in Setup 1 and Setup 2, respectively. }
    \label{local-l1-merged}
\end{figure}

\begin{figure}[H]
    \centering
    \includegraphics[width=1\linewidth]{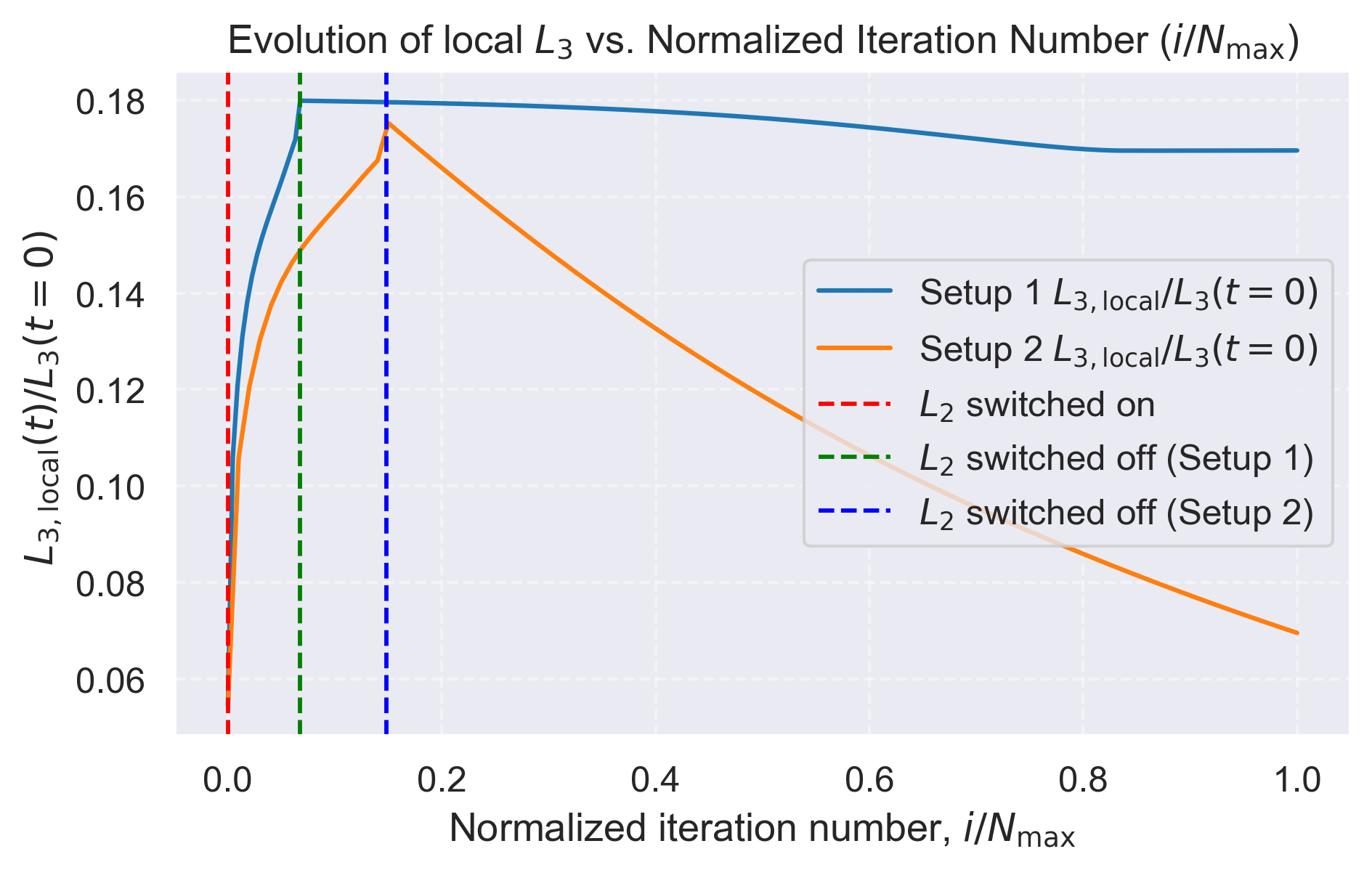}
    \caption{Evolution of the local metric $L_{3, \text{local}}$, normalized to its initial value, as a function of the normalized iteration number $i / N_{\text{max}}$ for Setup 1 (blue curve) and Setup 2 (orange curve). The vertical red dashed line indicates the iteration at which $L_2$ is switched on, while the green and blue dashed lines mark the iterations at which $L_2$ is switched off in Setup 1 and Setup 2, respectively. }
    \label{local-l3-merged}
\end{figure}

The contribution from volumes enclosing loop influence zones (regions closer to the loops), which we call "local" $L_1$ and $L_3$, is plotted in Figs. \ref{local-l1-merged} and \ref{local-l3-merged} for both cases, respectively. The purpose of such an evaluation is to better understand the impacts of the loops on the divergence and force-freeness on a more local scale since both $L_1$ and $L_3$ are global constraints.

In particular, the plots \ref{local-l1-merged} and \ref{local-l3-merged} better illustrate the initial disturbance (increase in $L_1$ and $L_3$) due to the loop inclusion, the monotonic decrease of $L_1$ (and of $L_3$ for setup II) after switching off the constraint $L_2$, the attenuation characteristic of reaching a minimum, stationary state (seen in Fig. \ref{local-l1-merged}), and the fact that the final $L_1$ and $L_3$ are, at a local scale closer to the loops, bounded by their initial values.

\end{appendix}

\end{document}